# Microwave fields driven domain wall motions in antiferromagnetic nanowires

Z. Y. Chen[1,*], Z. R. Yan[1,*], Y. L. Zhang[1], M. H. Qin[1,†], Z. Fan[1], X. B. Lu[1], X. S. Gao[1], and J. –M. Liu[2,†]

[1]*Institute for Advanced Materials, South China Academy of Advanced Optoelectronics and Guangdong Provincial Key Laboratory of Quantum Engineering and Quantum Materials, South China Normal University, Guangzhou 510006, China*

[2]*Laboratory of Solid State Microstructures and Innovative Center for Advanced Microstructures, Nanjing University, Nanjing 210093, China*

**[Abstract]** In this work, we study the microwave field driven antiferromagnetic domain wall motion in an antiferromagnetic nanowire, using the numerical calculations based on a classical Heisenberg spin model. We show that a proper combination of a static magnetic field plus an oscillating field perpendicular to the nanowire axis is sufficient to drive the domain wall propagation along the nanowire with the axial magnetic anisotropy. More importantly, the drift velocity at the resonance frequency is comparable to that induced by temperature gradients, suggesting that microwave field can be a very promising tool to control domain wall motions in antiferromagnetic nanostructures. Furthermore, the dependences of resonance frequency and drift velocity on the static and oscillating fields, the axial anisotropy, and the damping constant are discussed in details. This work provides useful information for the spin dynamics in antiferromagnetic nanostructures for spintronics applications.

[*]Z. Chen and Z. Yan contributed equally to the work.
[†]Authors to whom correspondence should be addressed. Electronic mail: qinmh@scnu.edu.cn and liujm@nju.edu.cn

## I. Introduction

Since the effective manipulation[1-4] and detection[5] of antiferromagnetic (AFM) states were realized, AFM materials have attracted extensive attentions due to their potential applications for AFM spintronics. In comparison with ferromagnets based storage devices,[6,7] antiferromagnets based devices are believed to hold several additional merits.[8] On the one hand, the AFM states are rather stable even under high magnetic fields, and information stored in AFM domains and/or domain walls (DWs) is also insensitive to applied magnetic fields. Furthermore, an AFM element would not magnetically disturb its neighbors due to its zero net magnetic moment, allowing the element arrangements in an ultrahigh density, noting that the demagnetization effect with AFM structure is essentially minimized. On the other hand, due to additional spin wave (SW) modes with higher frequencies in AFM structures, spin dynamics in antiferromagnets can be extremely fast. These advantages thus favor AFM materials for promising potentials in future devices.

However, one of the major challenges for AFM spintronic applications, at least on the current stage, is how to manipulate the AFM domain walls and their motion via a spin dynamics scenario, noting that the magnetic field and spin-polarized current may not work for an effective manipulation of AFM domain wall (DW) propagation due to the absence of net magnetic moment of the AFM domains. Based on the knowledge on the spin dynamics in ferromagnets,[9-12] various efforts in searching for control scheme for the AFM DW motion have been reported.[13-20] For example, a high AFM DW velocity may be achieved if the so-called Néel spin-orbit torque is utilized, as predicted theoretically[13] and then experimentally observed in CuMnAs.[21,22] This progress allows CuMnAs to be a good candidate for AFM spintronic applications. Moreover, it has been revealed that several SW modes can be used to drive the AFM DW motions, e.g. linear and circular SW can drive the DWs to move towards and away from the source, respectively.[15] Most recently, thermally driven AFM DW motion under a temperature gradient has been predicted by numerical calculations, where the competition between the entropic torque and the Brownian force is suggested to drive the AFM DW motion.[16-18] More interestingly, in these control schemes, the AFM DW remains non-tilted, indicating the higher wall mobility due to the missing of the walker breakdown.[17]

While the temperature gradient driven AFM DW motion can be a scheme with more theoretical sense rather than realistic sense, electrically driven scheme is certainly more attractive. It has been evidenced that static magnetic field or electric current driving is an ineffective scheme for the AFM DW dynamics. Alternatively, the Néel spin-orbit torque driving as a possible scheme may be given sufficient attention. Along this line, a proper microwave field can be a favored choice, noting that the microwave driven DW motion in ferromagnets has been demonstrated.[12] In fact, earlier preliminary investigation suggested that polarized microwave fields[19] or asymmetric field pulses[20] do work for exciting the dynamics of an AFM DW. For example, as early as 1994, the collinear AFM DW motion driven by an oscillating magnetic field was predicted, based on the perturbation theory, and the dependence of the drift velocity on the frequency and polarization of the field was discussed.[19]

However, only the case of oscillating field with small amplitude could be studied by the perturbation theory, and the drift velocity of the DW is estimated to be ~ 1.0 cm/s for a field amplitude of 10 Oe, which is too low to be used for spintronics devices.[19] Interestingly, it has been reported that the drift velocity of DW in ferromagnetic nanostructure is estimated to be ~ 20 m/s, induced by microwave field with amplitude of 100 Oe at the critical frequency, demonstrating the prominent effect of field amplitude on the drift velocity.[12] Thus, it is expected that an oscillating magnetic field with a large amplitude probably leads to a very fast AFM DW motion,[20] although its effects on the AFM dynamics are still not clear. Furthermore, the microwave driven DW motion has not been confirmed directly by numerical calculations or experiments, as far as we know. Thus, considering the limitation of current technical ability in experiments which prohibits a complete realization of theoretical prediction, the AFM DW motion driven by microwave field urgently deserves for investigation numerically in order to uncover the physical mechanisms of AFM dynamics and promote the application process for AFM spintronics.

In this work, we study the microwave field driven AFM DW motion in an AFM nanowire, based on the classical Heisenberg spin model. We figure out that high velocity of the DW motion is achieved with a proper combination of a static magnetic field plus an oscillating field at the resonance frequency. Moreover, the dependences of resonance frequency and drift velocity on the static and oscillating fields, the axial anisotropy, and the damping constant are discussed in details.

The remaining part of this paper is organized as follows: Sec. II is attributed to model and method. In Sec. III, the simulation results and discussion are presented and we make a conclusion in Sec. IV.

## II. Model and method

We numerically study the microwave field driven AFM DW motion based on the classical Heisenberg spin model with the isotropic nearest neighbors exchange interactions and biaxial anisotropy.[17,23] The model Hamiltonian is given by

$$H = -J\sum_{<i,j>} \mathbf{S}_i \cdot \mathbf{S}_j - d_x\sum_i (S_i^x)^2 - d_z\sum_i (S_i^z)^2 - \sum_i \mathbf{h} \cdot S_i^y, \tag{1}$$

where $J < 0$ is the AFM coupling constant, $d_z > d_x > 0$ are the anisotropic constants defining an easy axis in the $z$ direction and an intermediate axis in the $x$ direction, $\mathbf{S}_i = \boldsymbol{\mu}_i/\mu_s$ represents the normalized magnetic moment at site $i$ with the three components $S_i^x$, $S_i^y$ and $S_i^z$.

We consider an AFM nanowire with axis along the $z$ direction. The microwave field $h = h_{ac}\cdot\sin\omega t + h_{dc}$ is applied along the $y$ direction with a static field $h_{dc}$ and an oscillating field with frequency $\omega$ and amplitude $h_{ac}$. The AFM dynamics at zero temperature is studied by the Landau-Lifshitz-Gilbert (LLG) equation,[24,25]

$$\frac{\partial \mathbf{S}_i}{\partial t} = -\frac{\gamma}{\mu_s(1+\alpha^2)}\mathbf{S}_i \times [\mathbf{H}_i + \alpha(\mathbf{S}_i \times \mathbf{H}_i)], \tag{2}$$

where $\gamma$ is the gyromagnetic ratio, $\mathbf{H}_i = -\partial H/\partial \mathbf{S}_i$ is the effective field, and $\alpha$ is the Gilbert damping constant.

The nanowire is defined by a $2 \times 2 \times 181$ elongated three dimensional lattice (lattice parameter $a$), and the LLG simulations are performed using the fourth-order Runge-Kutta method with time step $\Delta t = 8.0 \times 10^{-5}$ $\mu_s/\gamma J$ (here, $J = |J|$ is the energy unit). Here, the periodic boundary conditions are applied in the $x$, $y$, and $z$ directions to reduce the finite-size effect, and the main conclusion will not be affected by such a choice. Unless stated otherwise, $d_x = 0.004J$, $d_z = 0.1J$, $\alpha = 0.01$, and $\gamma = 1$ are chosen in our simulations. After sufficient relaxation of the Néel AFM DW (depicted in Fig. 1), we apply microwave fields and study the DW motion. The staggered magnetization $2\mathbf{n} = \mathbf{m}_1 - \mathbf{m}_2$ is calculated to describe the spin dynamics, where $\mathbf{m}_1$ and $\mathbf{m}_2$ are the magnetizations of the two sublattices (occupied with the red and blue spins, respectively).[17]

## III. Simulation results and discussion

Fig. 2(a) gives the position of an AFM DW under a static magnetic field $h_{dc} = 0.2$. Unlike the FM DW which propagates along a wire under nonzero $h_{dc}$,[10] the AFM DW could not be efficiently driven by the static field. Specifically, the DW quickly shifts and stays at a new equilibrium position ($z = 84a$) when $h_{dc}$ is applied, and then returns to its original position as soon as the field is turned off at $\tau = 300$. Similarly, the AFM DW could not either be driven by an oscillating magnetic field alone ($h_{dc} = 0$). The DW position as a function of $\tau$ for field amplitude $h_{ac}$ = 0.4 at $\omega$ = 0.35 is presented in Fig. 2(b), as an example. It is clearly shown that the DW under a nonzero $h_{ac}$ only oscillates around its equilibrium position and cannot propagate along the wire. Interestingly, a combination of $h_{dc}$ plus $h_{ac}$ may result in an oscillatory propagation of the AFM DW. For example, in addition to the oscillatory motion, a steady propagation of the DW along the wire at a constant drift velocity is observed for $h_{dc}$ = 0.2 and $h_{ac}$ = 0.4 at $\omega$ = 0.35, as clearly shown in Fig. 2(c). More importantly, the drift velocity is significantly dependent of frequency $\omega$, and a high DW mobility could be achieved at particular frequencies. Fig. 2(d) gives the results for $\omega$ = 1, which clearly demonstrates the DW propagation with a high drift velocity.[19]

The simulated results can be qualitatively understood from the competition between various torques acting on the DW.[17] Following Eq. (2), a magnetic field along the $y$ direction leads to two types of torques on the central plane of an AFM DW, as depicted in Fig. 1. In details, the first torque ~ $-\mathbf{S} \times \mathbf{H}$ pointing in an opposite direction on the two sublattices and drives the DW, while the second torque ~ $-\mathbf{S} \times (\mathbf{S} \times \mathbf{H})$ pointing in the same direction would not tilt the wall but slightly drive the spins out of their easy plane. When a static magnetic field is applied, additional exchange and anisotropic fields are induced by the deviation of the spins from the easy plane, resulting in additional torque opposite to that from the external field (black arrow in Fig. 1). As a result, the first torque ~ $-\mathbf{S} \times \mathbf{H}$ is quickly reduced to zero, and the DW stops at a new equilibrium position. Furthermore, the DW returns to its original equilibrium position driven by the first torque when the static field is decreased to zero, as confirmed in our simulations.

When an oscillating magnetic field is considered, the exchange and anisotropic fields induced by the spin deviation always fall behind the applied field. At $h_{dc}$ = 0, the first torque is antisymmetric and changes its direction during one oscillation period, resulting in the DW oscillation. Thus, the DW could not be efficiently driven to propagate along the wire. Importantly, the symmetry of the first torque (~ $-\mathbf{S} \times \mathbf{H}$, nonlinearly changes with the applied magnetic field) is broken by further considering the nonzero $h_{dc}$, and a net driven torque during a period is available. As a result, the DW could be efficiently driven by the net torque resulted from a proper combination of the static and oscillating magnetic fields, as demonstrated in our simulations.

The drift velocity of the AFM DW, $v$, is significantly dependent of $\omega$. As a matter of fact, earlier work demonstrated the high DW mobility when an oscillating field matches the characteristic frequencies of the magnetic domains.[22] Fig. 3 gives the calculated $v$ as a function of $\omega$ at $h_{ac}$ = 0.4 and $h_{dc}$ = 0.2, which shows three $v$ maxima at frequencies $\omega$ = 0.35,

0.76, and 1.02, respectively. Especially, the highest velocity $v_{max}$ is obtained at the resonance frequency $\omega_0 = 1.02$, which is comparable to that of the thermally driven AFM DW motion. Moreover, the DW may move toward opposite direction (at $\omega = 1.03$, for example) by elaborately modulating the frequency of the applied field, allowing one to control DW motion at ease, which is particularly meaningful for future applications.[19] Thus, our work strongly suggests that microwave fields can be another control parameter for AFM DW motion in AFM nanostructures.

Subsequently, we investigate the dependences of resonance frequency $\omega_0$ and highest drift velocity $v_{max}$ on various factors. Fig. 4(a) gives the simulated $v(\omega)$ curves for various $h_{ac}$ for $h_{dc} = 0.2$. It is clearly shown that $\omega_0$ is less affected by $h_{ac}$, while $v_{max}$ increases with the increasing $h_{ac}$. Thus, it is indicated that $\omega_0$ is mainly related to the characteristic frequency of the magnetic domain which is hardly affected by $h_{ac}$. Furthermore, the net torque in one oscillation period is enlarged as $h_{ac}$ increases, speeding up the DW at $\omega_0$. Moreover, our work also reveals that both $v_{max}$ and $\omega_0$ can be modulated by $h_{dc}$, as shown in Fig. 4(b) which presents the $v(\omega)$ curves for various $h_{dc}$ at $h_{ac} = 0.4$. On the one hand, higher $h_{dc}$ increases the spin precession angular velocity, and in turn results in a higher resonance frequency. Thus, $\omega_0$ increases with the increase of $h_{dc}$, qualitatively consistent with the earlier theoretical calculations, $\omega_0 \sim \gamma(2H_EH_A + h_{dc}^2)^{1/2}$ where $H_A$ and $H_E$ are respectively the anisotropy field and the exchange field.[26,27] On the other hand, the net driven torque in one period is also enlarged, leading to the increase of $v_{max}$ and the enlargement of the region of $\omega$ at which the AFM DW can be efficiently driven.

In Fig. 5(a), we give the simulated $v(\omega)$ curves for various $d_x$ at $h_{ac} = 0.4$, $h_{dc} = 0.2$ and $d_z = 0.1$. It is demonstrated that both $v_{max}$ and $\omega_0$ are increased as $d_x$ increases. When only the biaxial anisotropy is considered, the precession equations of a single spin $\mathbf{S}$ are[27]

$$\begin{cases} S_x' = -\gamma d_z S_y S_z \\ S_y' = -\gamma (d_x - d_z) S_x S_z \\ S_z' = \gamma d_x S_x S_y \end{cases}, \qquad (3)$$

where $S_k' = \partial S_k / \partial t$ ($k = x, y, z$). Considering the fact that both $S_y$ and $S_y'$ are very small due to the strong biaxial anisotropy, one can obtain the characteristic frequency of the spin precession $\omega \sim \gamma S_y (d_x d_z)^{1/2}$. As a result, the resonance frequency of the DW motion increases when the anisotropy is enhanced, as further confirmed in Fig. 5(b) which presents the calculated $v(\omega)$ curves for various $d_z$ at $d_x = 0.004$. However, converse to the case of $d_x$, larger $d_z$ results in a smaller $v_{max}$, which can be qualitatively understood from the energy landscape. Based on the

continuum model, the total energy of the DW can be expressed as[28]

$$E_{DW} = N_D \cdot 2\sqrt{2(d_z - d_x)J} \,, \tag{4}$$

where $N_D$ is the number of spins in the cross section along the $z$ axis. It is well noted that $E_{DW}$ (its value depends on $d_z - d_x$) determines the mobility of the DW, specifically, higher $E_{DW}$ leads to a lower mobility of the DW. As a result, $v_{max}$ is increased with the increase/decrease of $d_x/d_z$, as demonstrated in our simulations.

At last, the effect of the damping constant $\alpha$ on $v_{max}$ and $\omega_0$ is also investigated, and the simulated results are shown in Fig. 5(c). It is clearly shown that both $v_{max}$ and $\omega_0$ are decreased with the increase of $\alpha$, demonstrating the fact that an enhanced damping term lowers the mobility of the DW. On the one hand, the damping term always impedes the spin precession and diminishes the characteristic frequency of magnetic domain, resulting in the decreases of $\omega_0$ and the effectively driven region of $\omega$. On the other hand, the second torque ~ $-\mathbf{S} \times (\mathbf{S} \times \mathbf{H})$ is significantly enhanced with the increasing $\alpha$, and arranges the spins along the effective field $\mathbf{H}$ direction more quickly. As a result, the first torque ~ $-\mathbf{S} \times \mathbf{H}$ during an oscillating period is reduced, which speeds down the DW.

It is easily noted that an in-plane magnetic field which can efficiently drive the DW motion in ferromagnets through the torque ~ $-\mathbf{S} \times (\mathbf{S} \times \mathbf{H})$ could not drive the motion of the AFM DW. Specifically, for $\mathbf{h}$ along the $z$ direction, the torque ~ $-\mathbf{S} \times (\mathbf{S} \times \mathbf{H})$ points into the same direction on the two sublattices, and these two contributions cancel in antiferromagnets. Thus, AFM DW could not be efficiently driven by in-plane microwave fields. This phenomenon has been confirmed in our simulations, although the corresponding results are not shown here for brevity.

## IV. Conclusion

In general, we have studied the AFM DW motion driven by microwave fields using the LLG simulations of the classical Heisenberg spin model. It is revealed that a proper combination of a static field and an oscillating field can drive the DW motion in AFM nanostructures, and a large drift velocity comparable to that induced by temperature gradients can be obtained at the resonance frequency. Furthermore, the dependences of the resonance frequency and velocity on several parameters are investigated and explained in details. Thus, our work provides useful information for the spin dynamics in AFM nanostructures for spintronics applications.

**Acknowledgement:**

The work is supported by the National Key Projects for Basic Research of China (Grant No. 2015CB921202), and the National Key Research Program of China (Grant No. 2016YFA0300101), and the Natural Science Foundation of China (Grant Nos. 51332007, 51431006), and the Science and Technology Planning Project of Guangdong Province (Grant No. 2015B090927006). X. Lu also thanks for the support from the project for Guangdong Province Universities and Colleges Pearl River Scholar Funded Scheme (2016).

**FIGURE CAPTIONS**

Fig.1. (color online) Draft of the torques acting on the central plane of a DW under a microwave field. Blue and red arrows represent spins in two sublattices.

Fig.2. (color online) The domain wall position as a function of time $\tau$ for (a) $h_{dc} = 0.2$ for $\tau < 300$ and $h_{dc} = 0$ for $\tau > 300$ at $h_{ac} = 0$, and (b) $h_{ac} = 0.4$ at $\omega = 0.35$ and $h_{dc} = 0$, and (c) $h_{ac} = 0.4$ at $\omega = 0.35$ and $h_{dc} = 0.2$, and (d) and $h_{ac} = 0.4$ at $\omega = 1.0$ and $h_{dc} = 0.2$.

Fig.3. (color online) The calculated DW drift velocity $v$ as a function of the oscillating frequency $\omega$ for $h_{ac} = 0.4$ and $h_{dc} = 0.2$.

Fig.4. (color online) The calculated $v(\omega)$ curves (a) for various $h_{ac}$ at $h_{dc} = 0.2$, and (b) for various $h_{dc}$ at $h_{ac} = 0.4$.

Fig.5. (color online) The calculated $v(\omega)$ curves (a) for various $d_x$ at $d_z = 0.1$, and (b) for various $d_z$ at $d_x = 0.004$, and (c) for various $\alpha$.

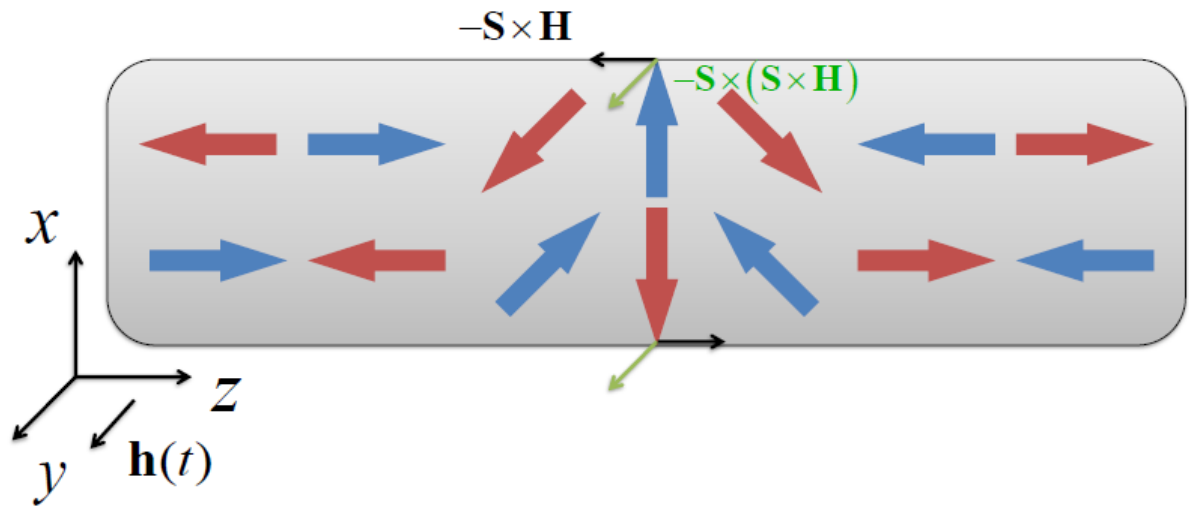

Fig.1. (color online) Draft of the torques acting on the central plane of a DW under a microwave field. Blue and red arrows represent spins in two sublattices.

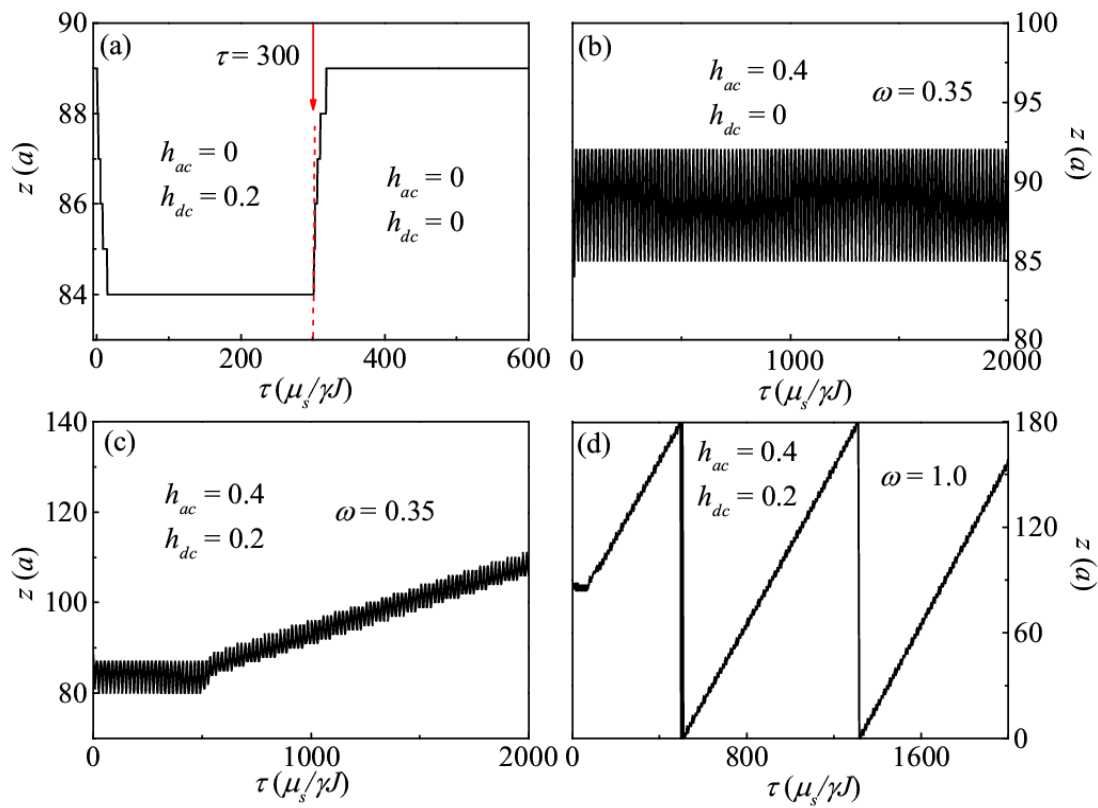

Fig.2. (color online) The domain wall position as a function of time $\tau$ for (a) $h_{dc} = 0.2$ for $\tau < 300$ and $h_{dc} = 0$ for $\tau > 300$ at $h_{ac} = 0$, and (b) $h_{ac} = 0.4$ at $\omega = 0.35$ and $h_{dc} = 0$, and (c) $h_{ac} = 0.4$ at $\omega = 0.35$ and $h_{dc} = 0.2$, and (d) and $h_{ac} = 0.4$ at $\omega = 1.0$ and $h_{dc} = 0.2$.

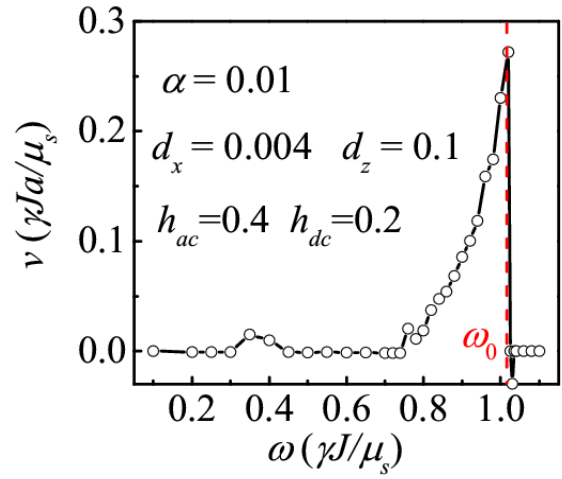

Fig.3. (color online) The calculated DW drift velocity $v$ as a function of the oscillating frequency $\omega$ for $h_{ac} = 0.4$ and $h_{dc} = 0.2$.

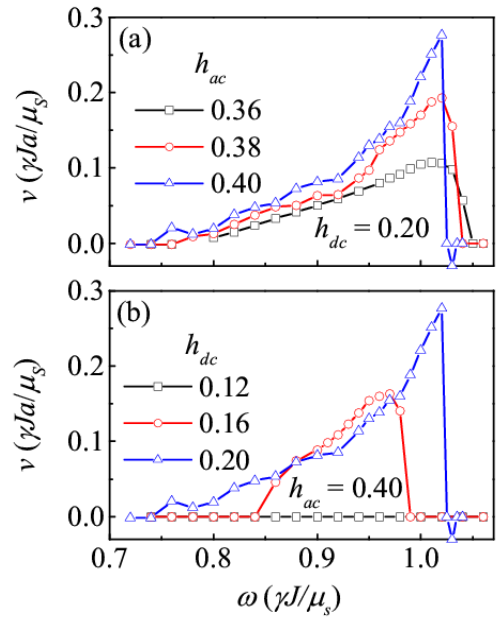

Fig.4. (color online) The calculated $v(\omega)$ curves (a) for various $h_{ac}$ at $h_{dc} = 0.2$, and (b) for various $h_{dc}$ at $h_{ac} = 0.4$.

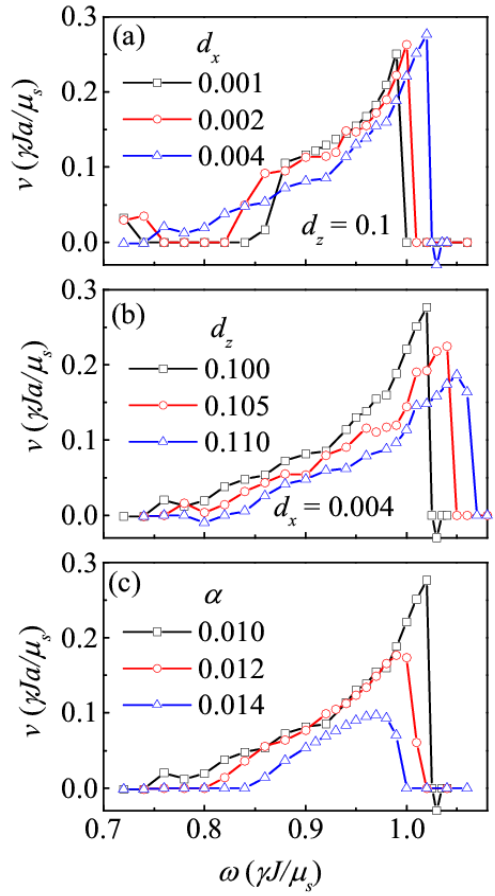

Fig.5. (color online) The calculated $v(\omega)$ curves (a) for various $d_x$ at $d_z = 0.1$, and (b) for various $d_z$ at $d_x = 0.004$, and (c) for various $\alpha$.